\documentclass{appolb}
\usepackage{epsfig}

\begin{document}

\title{Modelling CC neutrino cross sections in the few GeV energy region%
\thanks{Presented by J.T. Sobczyk at the {\sl Cracow Epiphany Conference on Neutrinos and Dark Matter},
Jan. 5-8, 2006, Cracow, Poland}%
}

\author{
Jaros\l aw A. Nowak
\and Jan T. Sobczyk
\address{Institute of Theoretical Physics, University of Wroc\l aw}
} \maketitle
\begin{abstract}
Selected problems in modelling neutrino-nucleon and -nuclei cross
sections in the neutrino energy region of the few GeV are
reviewed.

\end{abstract}

\section{Introduction}

The aim of this paper is to review some of recent developments in modelling the Charge Current (CC)
neutrino interactions with both free nucleons and nuclei targets in the few GeV neutrino energy
region \cite{NuInt}. This energy range is characteristic for atmospheric neutrinos and for several
running or approved long-baseline experiments. The knowledge of the cross sections is necessary for
future more precise measurements of neutrino oscillation parameters. In our discussion we adopt a
practical approach and will always have in mind Monte Carlo (MC) implementation of presented
models.

The few GeV energy region is rather complicated because three different dynamical formalisms are
relevant: of quasi-elastic reactions, resonance excitation and more inelastic channels treated
together in the DIS formalism. The significance of three dynamics is seen in Fig. 1 where the total
CC cross sections for the muon neutrino scattering off free {\it isoscalar} nucleon target (i.e.
average from proton and neutron targets)is presented. The contributions from quasi-elastic, single
pion production (SPP) and more inelastic channels (denoted as DIS) are also shown separately. It is
seen that in the few GeV energy region all three contributions are important. The  cross section
from SPP channels in Fig. 1 is restricted by the condition on the invariant hadronic mass
$W<W_{cut}=2$~GeV.

%%%%%%%%%%%%%%%%%%%%%%%%%%%%%%%%%%%%%%%%%%%%%%%%%%%%%%%%%%%%%%%%%%%%%%%%%%%
\begin{figure}\label{rys_total_nu_cc}
  \includegraphics[scale=0.9]{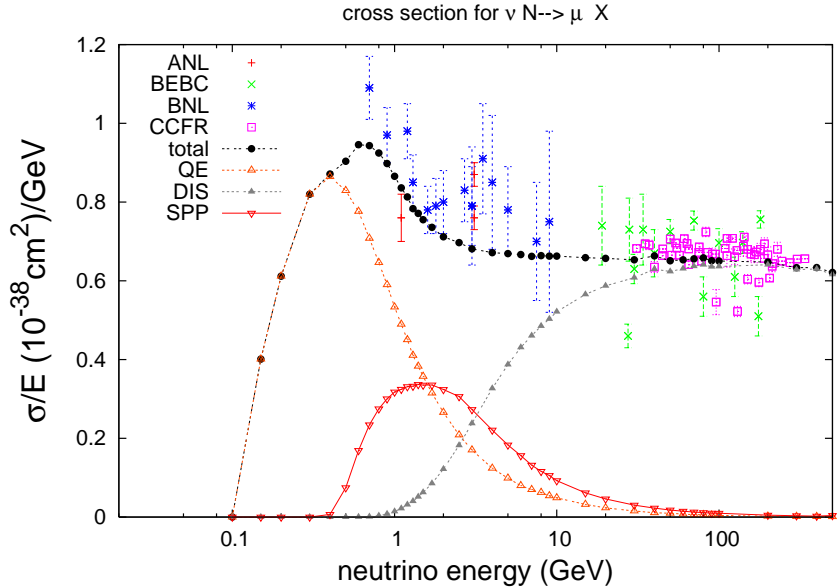}\\
  \caption{Total cross section for $\nu_\mu$ CC scattering on isoscalar target as predicted by the WROCLAW MC
  generator.
  The contributions from quasi-elastic, single pion production (SPP) and more inelastic channels (denoted as DIS) are
  also shown separately. }\label{rys_diff_q2_qel}
\end{figure}
%%%%%%%%%%%%%%%%%%%%%%%%%%%%%%%%%%%%%%%%%%%%%%%%%%%%%%%%%%%%%%%%%%%%%%%%%

%%%%%%%%%%%%%%%%%%%%%%%%%%%%%%%%%%%%%%%%%%%%%%%%%%%%%%%%%%%%%%%%%%%%%%%%%%%%
\begin{figure}
  \includegraphics[scale=0.9]{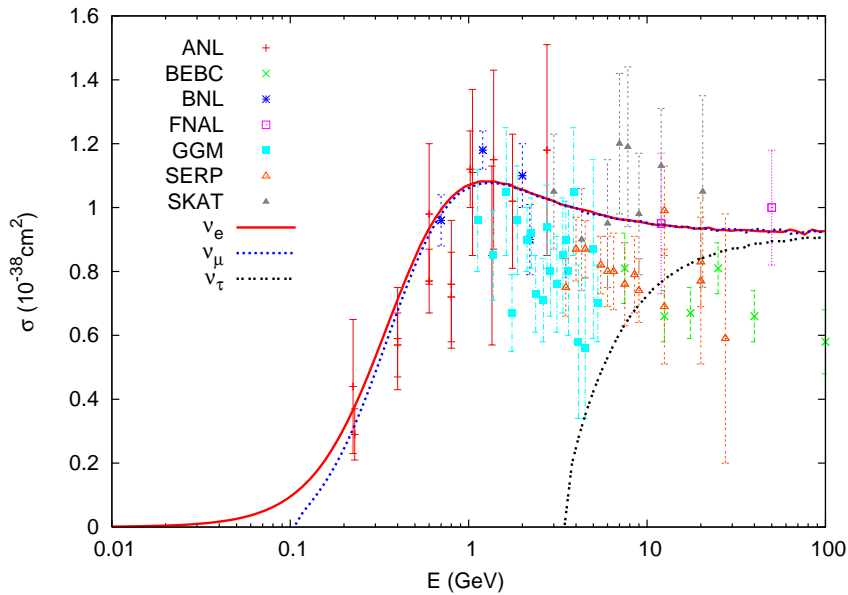}\\
  \caption{Quasi-elastic cross section for $\nu_e$, $\nu_\mu$ and $\nu_\tau$.
  Experimental points refer to $\nu_\mu$ scattering.}\label{rys_total_all}
\end{figure}
%%%%%%%%%%%%%%%%%%%%%%%%%%%%%%%%%%%%%%%%%%%%%%%%%%%%%%%%%%%%%%%%%%%%%%%%%%%%%

The plan of the paper is the following. We start in Section 2 from a description of quasi-elastic
reaction.  In Sections  3 and 4 we review models of SPP and the formalism of Deep Inelastic
Scattering. Sections 5 and 6 deal with nuclear effects.

\section{Quasi-elastic reactions}

There are two CC $\Delta S=0$ quasi-elastic channels: $\nu_l+n\rightarrow l^-+p$ and
$\bar\nu_l+p\rightarrow l^++n$. In the discussed energy region the condition $Q^2<<M_W^2$ holds and
it is enough to consider processes in the effective Fermi theory approximation.

The matrix element contains leptonic part which is exactly known and the hadronic one, which cannot
be calculated from first principles. The hadronic current contains four form-factors, functions of
$Q^2$. Vector form-factors $F_{1,2}$ are determined (CVC) by their electromagnetic counterparts.
The form-factor $F_P$ can be expressed in terms of $F_A$ (PCAC) \cite{LS}.  In recent years an
improvement to MC codes was introduced by a replacement of old-fashioned dipole form-factors with
one of the available fits to experimental data \cite{BBBA}. $F_A$ is considered to be in the dipole
form with two parameters: $g_A$ determined by the $\beta$ decay and the axial mass $M_A$ which is
not exactly known. The value of $M_A$ determines the shape of ${d\sigma}/{dQ^2}$ and the overall
quasi-elastic cross section \cite{Artur}.

%%%%%%%%%%%%%%%%%%%%%%%%%%%%%%%%%%%%%%%%%%%%%%%%%%%%%%%%%%%%%%%%%%%%%%%%%%%%%%%
\begin{figure}
  \includegraphics[scale=0.7]{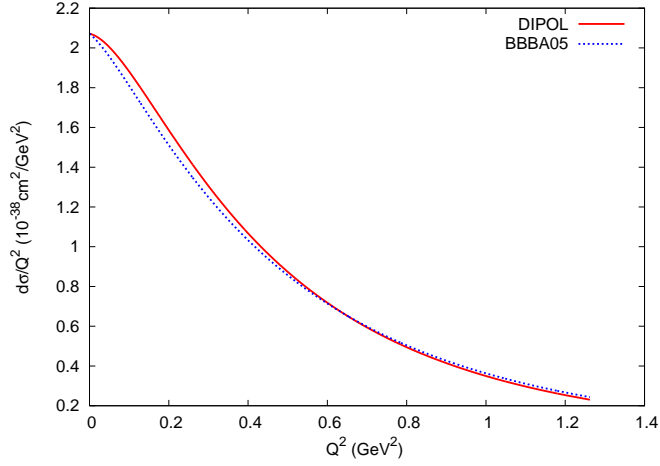}\\
  \caption{Modification of the shape of ${d\sigma}/{dQ^2}$ due to
non-dipole electromagnetic form factors. For both curves
$M_A=1.03$~GeV.
  }\label{rys_diff_q2_qel}
\end{figure}

\begin{figure}
  \includegraphics[scale=0.7]{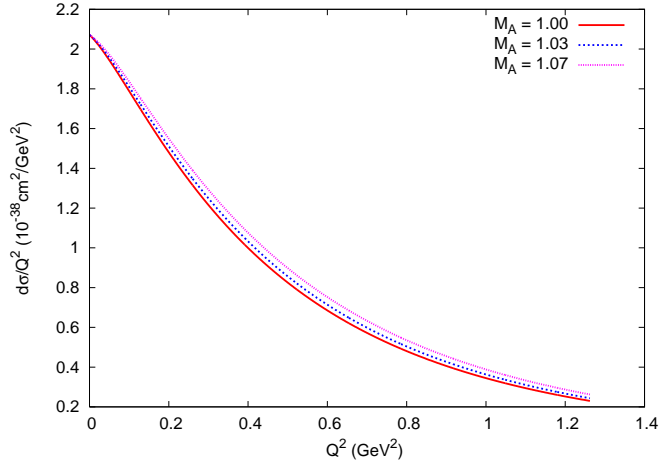}\\
  \caption{Modification of the shape of ${d\sigma}/{dQ^2}$ due to
different choices of the axial mass $M_A=1,\ 1.03,\ 1.07$~GeV. It
is seen that the choice of smaller $M_A$ reduces the overall cross
section in the different way than the substitution of dipole
form-factors by BBBA05 ones (see the Fig. 3).
  }\label{rys_diff_q2_qel}
\end{figure}
%%%%%%%%%%%%%%%%%%%%%%%%%%%%%%%%%%%%%%%%%%%%%%%%%%%%%%%%%%%%%%%%%%%%%%%%%%%%%%%%%%%

\section{Single pion production}

There are three CC SPP channels for neutrino reactions: $\nu_l+n\rightarrow l^-+p+\pi^0$,
$\nu_l+n\rightarrow l^-+n+\pi^+$, $\nu_l+p\rightarrow l^-+p+\pi^+$, and another three for
anti-neutrino reactions. The characteristic feature of SPP reactions is the appearance of the
$\Delta$ resonance in the differential cross section ${d\sigma}/{dW}$.

There are several theoretical models of SPP. Almost all Monte Carlo generators use the  Rein-Sehgal
(RS) model \cite{RS}. It includes contributions from 18 resonances of mass $M_{res}<2$~GeV treated
in the coherent way. The non-resonant contribution is then added in the incoherent way in order to
get an agreement with available data. An alternative model has been developed recently by the
Dortmund group \cite{Olga}. It is based on experimental data on electromagnetic helicity
amplitudes. The model contains the following resonances: $P_{33}(1232)$, $P_{11}(1440)$,
$D_{13}(1520)$ and $S_{11}(1535)$. It includes $m_l^2$ ($m_l$ is the charged lepton mass) terms
which reduce the ${d\sigma}/{dQ^2}$ by $\sim 5-10\%$ at low $Q^2$ and which are absent in the
original RS model. A similar model with few resonances was constructed many years ago by
Fogli-Nardulli \cite{FN}.

A common difficulty of SPP models is related to the issue of description of the non-resonant
background. The most systematic approach is the one adopted in the Sato-Lee model \cite{SL}. It is
based on the quark model with the pion cloud effects taken into account. It predicts the
non-resonant contribution to $\nu$-neutron SPP channels on the level of $\sim 25\%$ and to
$\nu$-proton SPP on the level of $\sim 5\%$. Alternative effective descriptions of non-resonant
background are introduced in some MC codes. The idea is to simulate the background by a fraction of
the DIS cross section in the resonance kinematical domain i.e. for $W<W_{cut}$ \cite{Wroclaw_old}
\cite{Gall}. Another theoretical possibility to deal with the non-resonant background is suggested
by the hypothetical two-component quark-hadron duality \cite{Harari}. If the hypothesis is true,
the background is given by the sea quark contribution to DIS structure functions. One can also try
to model the non-resonant background by assuming that its dependence on the invariant mass is
similar as in electron scattering i.e. $\sim\sqrt{W-W_{thr}}\sum a_j(Q^2)(W-W_{thr})^j$, where
$W_{thr}$ is the threshold for the pion production \cite{W_thr}.

The lack of precise experimental data makes it impossible to
select a preferred model of SPP.

%%%%%%%%%%%%%%%%%%%%%%%%%%%%%%%%%%%%%%%%%%%%%%%%%%%%%%%%%%%%%%%%%%%%%%%%%%%%%%%%%
\begin{figure}
  \includegraphics[scale=0.7]{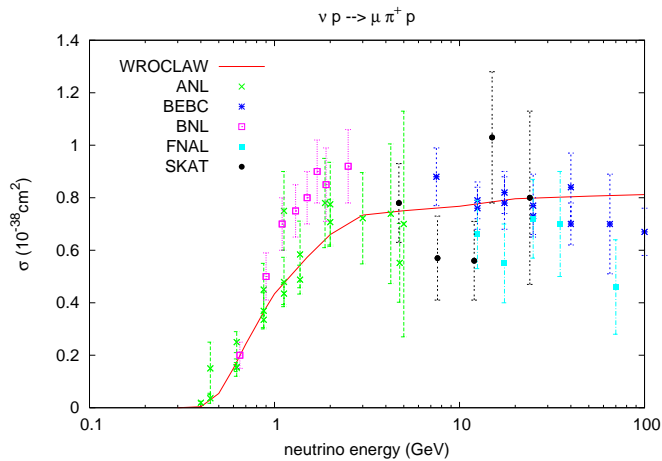}\\
  \caption{Cross section for $\nu_\mu p\rightarrow \mu^-p\pi^+$ as predicted by the WROCLAW MC
  generator.}
\end{figure}

\begin{figure}
  \includegraphics[scale=0.7]{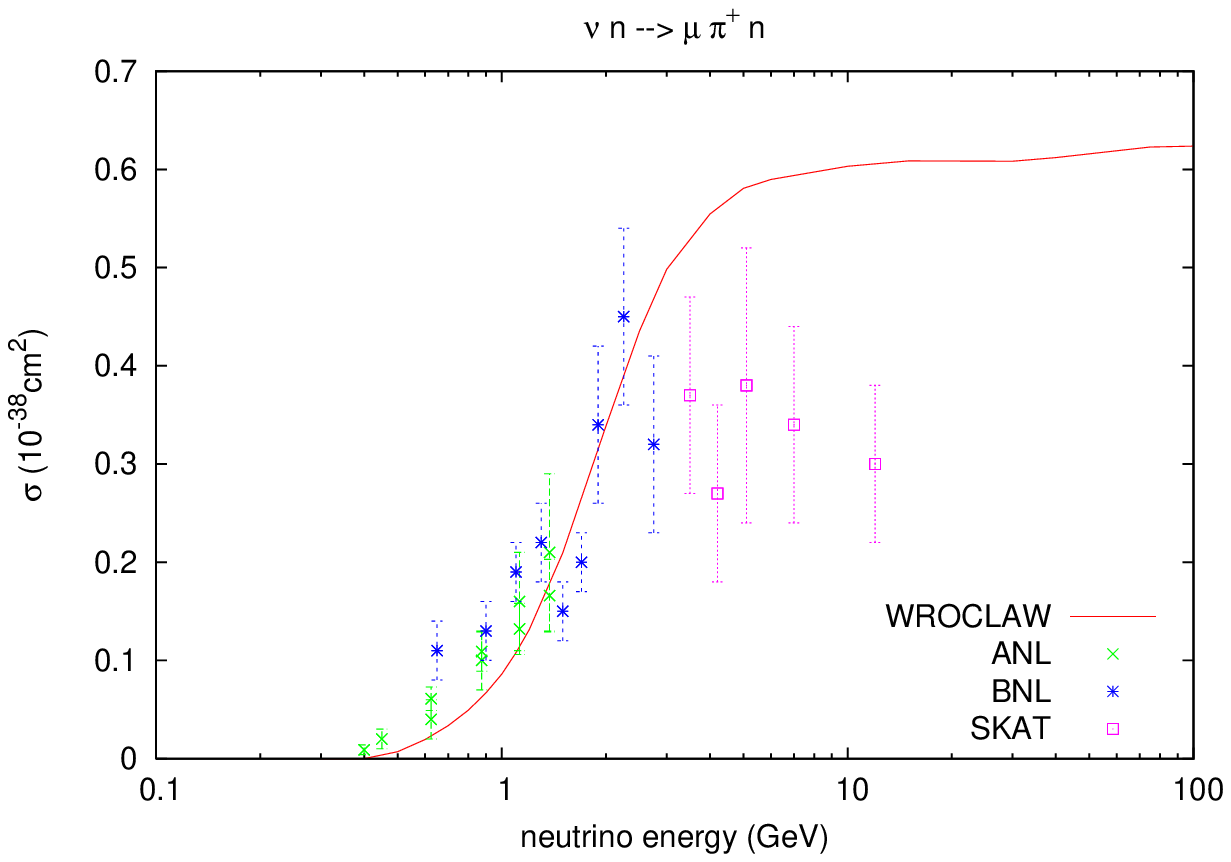}\\
  \caption{Cross section for $\nu_\mu n\rightarrow \mu^-n\pi^+$ as predicted by the WROCLAW MC
  generator.}
\end{figure}

\begin{figure}
  \includegraphics[scale=0.7]{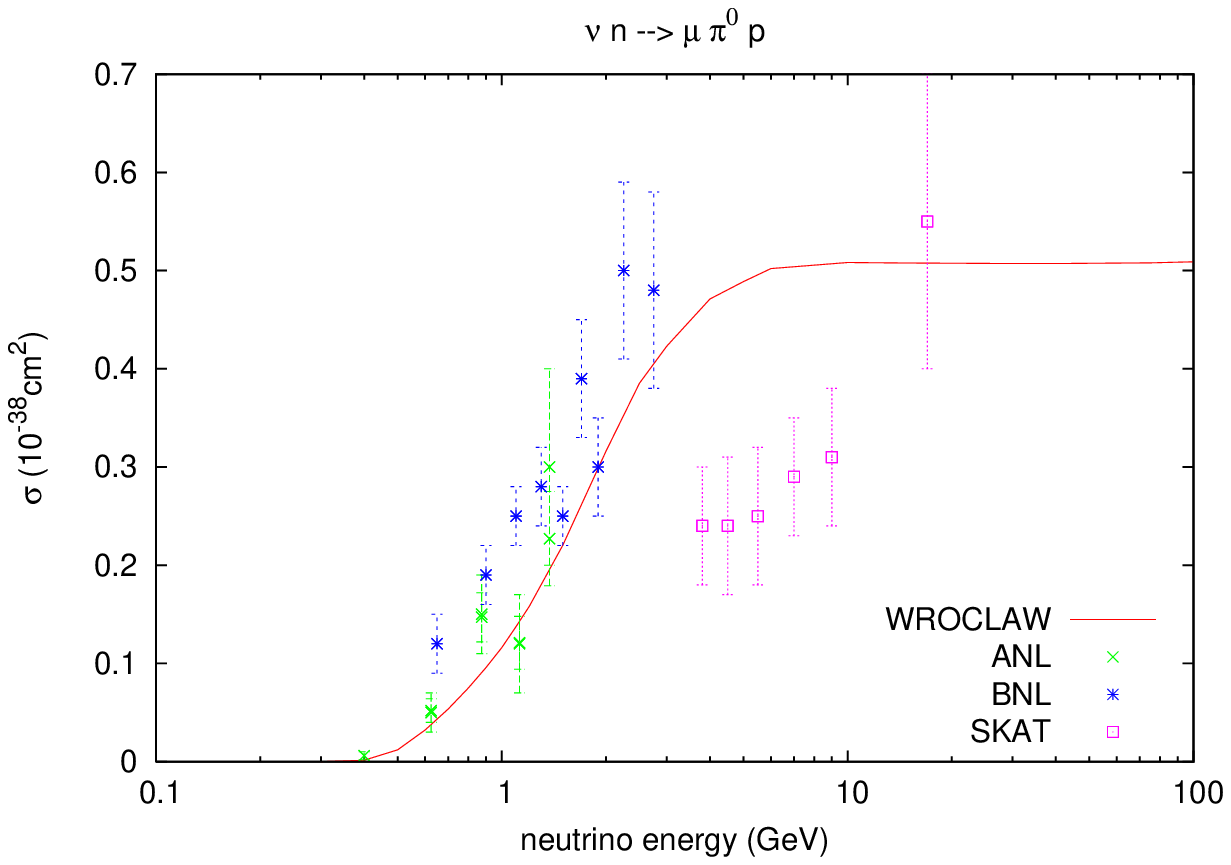}\\
  \caption{Cross section for $\nu_\mu n\rightarrow \mu^-p\pi^0$ as predicted by the WROCLAW MC
  generator.}
\end{figure}
%%%%%%%%%%%%%%%%%%%%%%%%%%%%%%%%%%%%%%%%%%%%%%%%%%%%%%%%%%%%%%%%%%%%%%%%%%%%%%%%%%%

%%%%%%%%%%%%%%%%%%%%%%%%%%%%%%%%%%%%%%%%%%%%%%%%%%%%%%%%%%%%%%%%%%%%%%%
\begin{figure}
\begin{tabular}{c c}
    \includegraphics[scale=0.3]{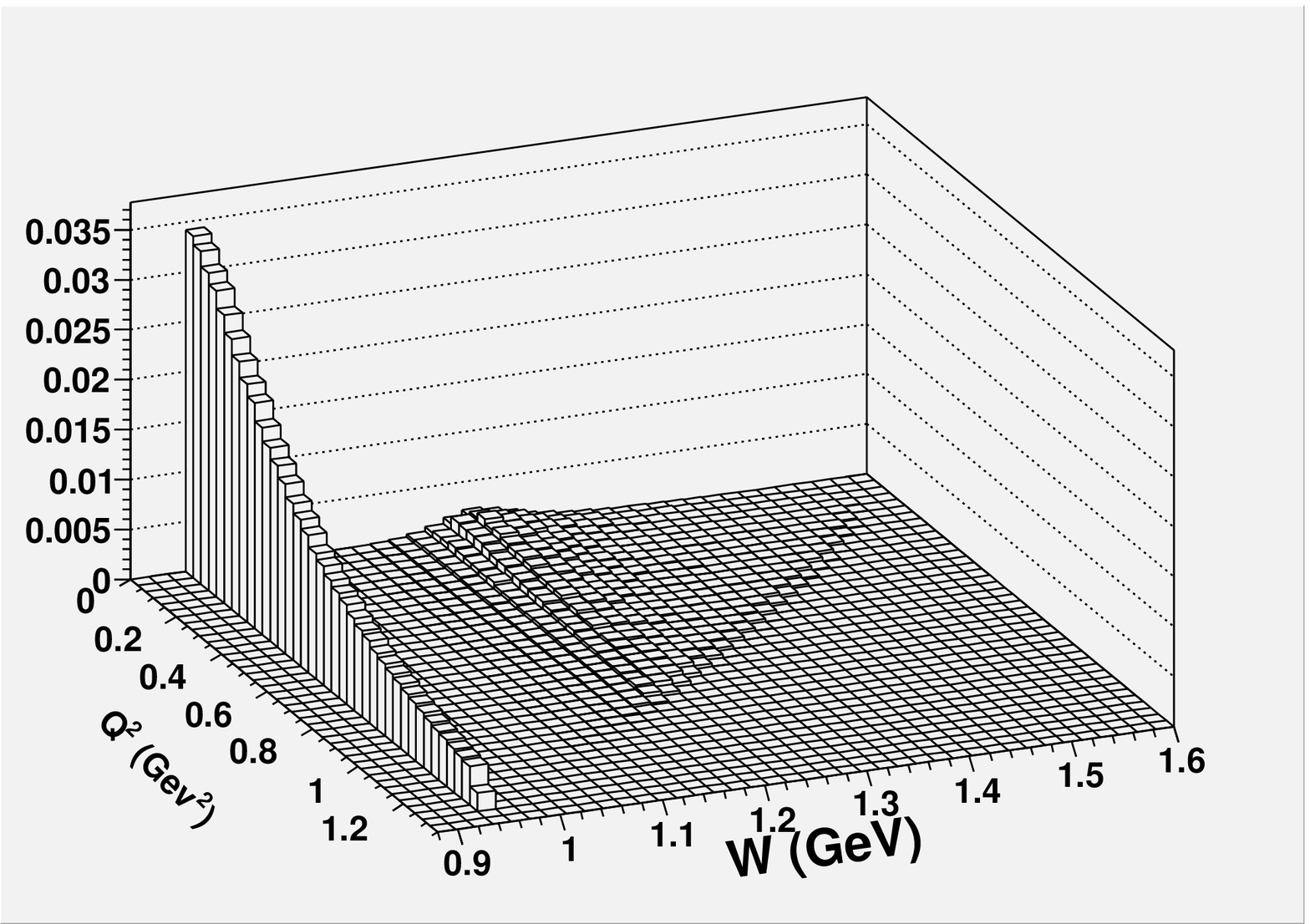}
 &   \includegraphics[scale=0.3]{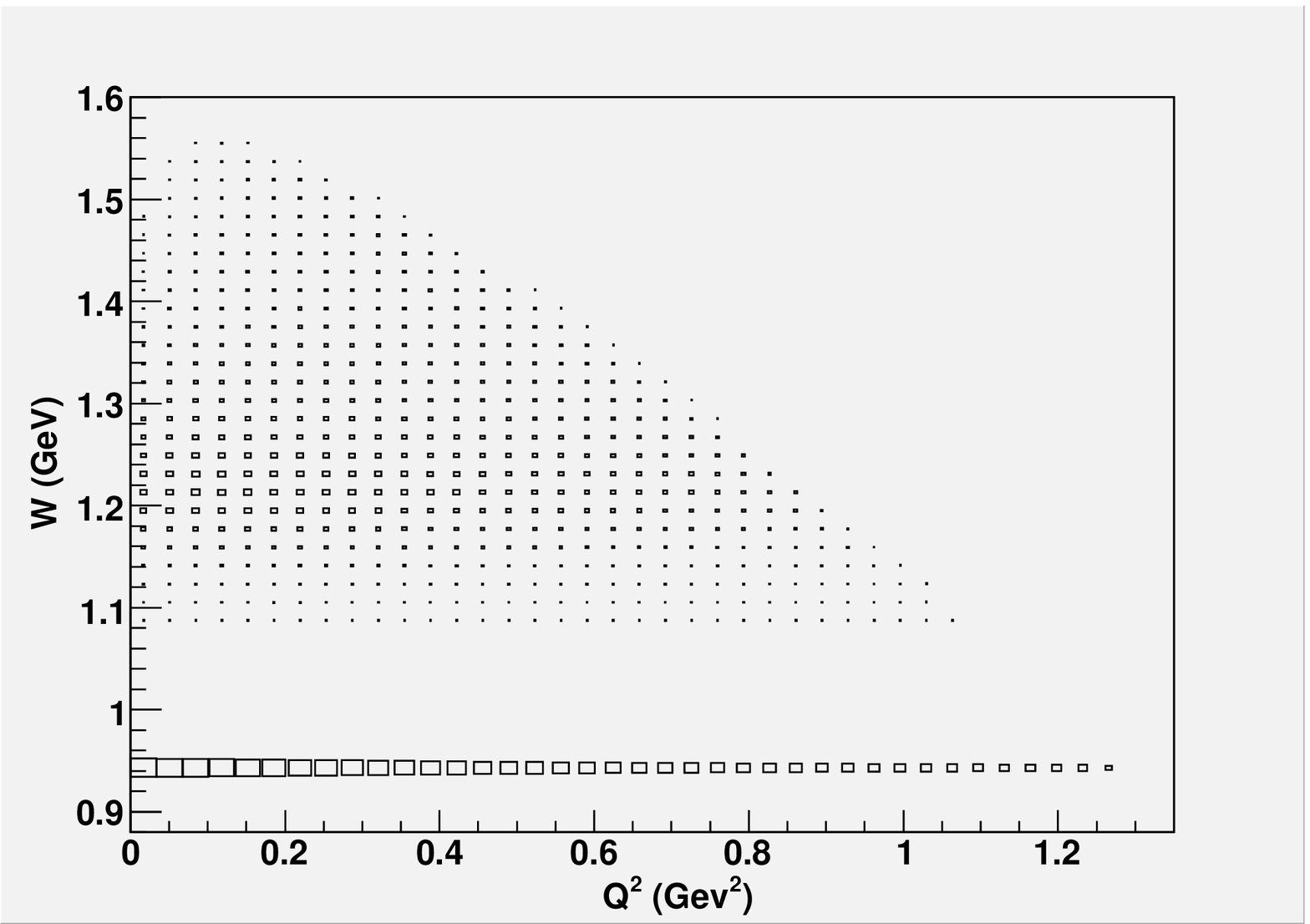}\\
    \includegraphics[scale=0.3]{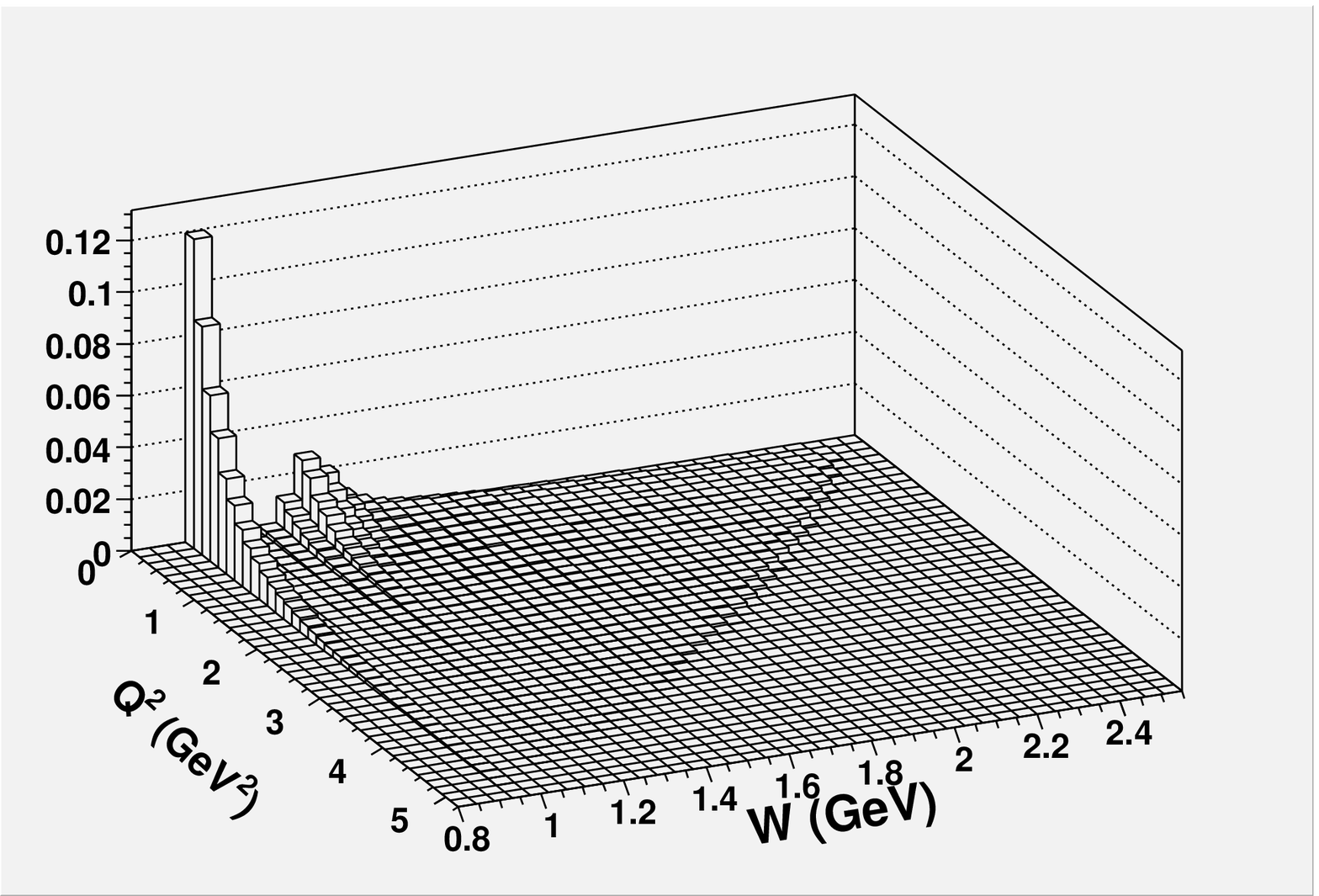}
 &   \includegraphics[scale=0.3]{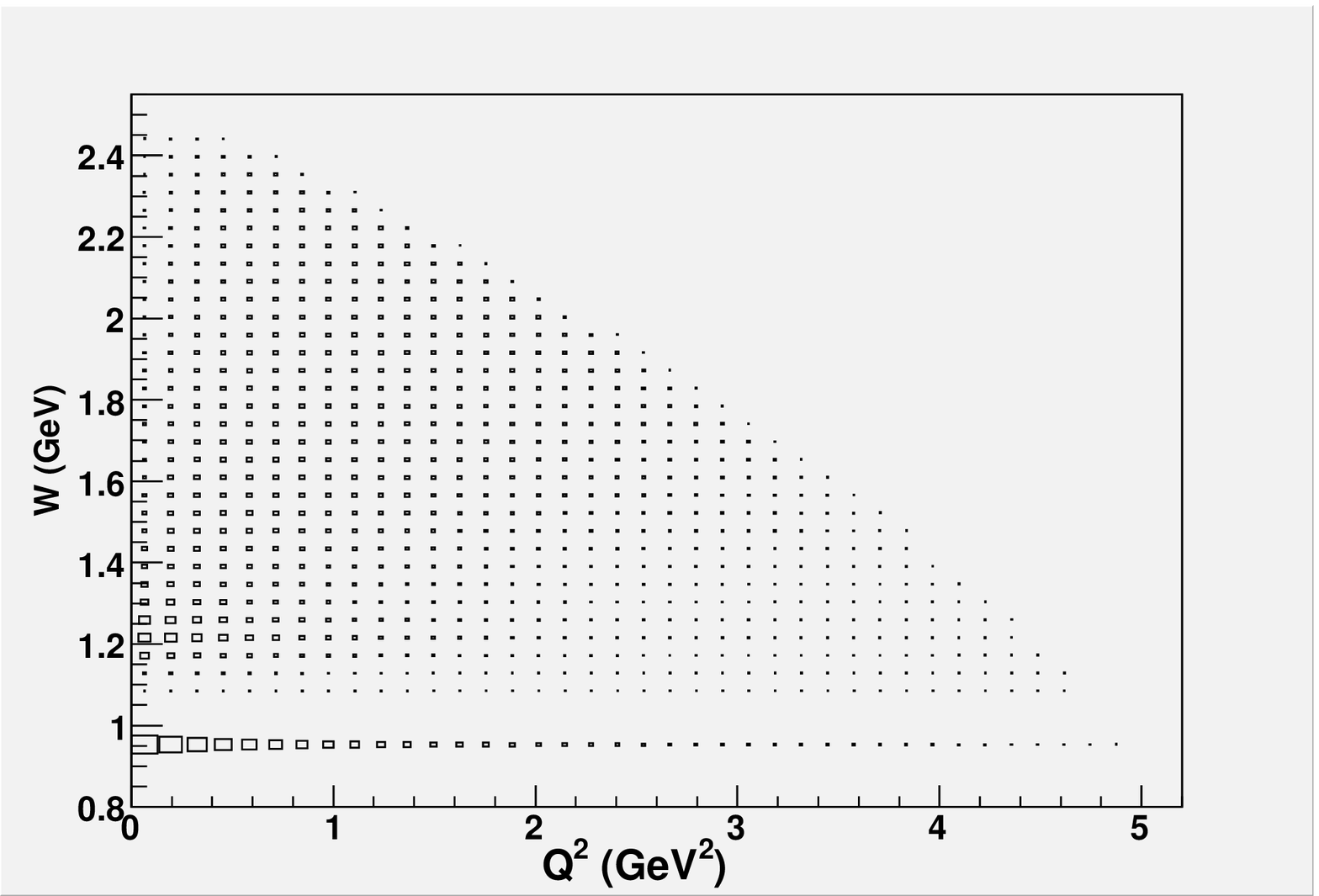}\\
\end{tabular}
  \caption{Distribution of events in $W$ and $Q^2$ for $\nu_\mu N$ scattering
  at
  energies $E_\nu=1$~GeV (top) and $E_\nu=3$~GeV (bottom) as predicted by the WROCLAW MC
  generator. The quasi-elastic contribution is seen the peak at $W=M$ (nucleon's mass). The
  $\Delta$ excitation region is also clearly seen. }
\end{figure}
%%%%%%%%%%%%%%%%%%%%%%%%%%%%%%%%%%%%%%%%%%%%%%%%%%%%%%%%%%%%%%%%%%%

\section{More inelastic channels}

The DIS formalism describes the inclusive $\nu$N cross section in the scaling limit. The
justification of the theory comes from the perturbative QCD. In the few GeV neutrino energy region
an important contribution to the cross section comes from the small $Q^2$ region, where the theory
behind the DIS formalism is not valid. Thus the first problem is to get a correct form of the
structure functions. The standard procedure is to express $F_{4,5}$ in terms of $F_{1,2}$ \cite{KR}
and then to express $F_1$ in terms of $F_2$ and $R\equiv({\sigma_L}/{\sigma_T})$. In the scaling
limit remaining $F_{2,3}$ are given as combinations of PDF's (parton distribution functions). In
the region we focus on, the target mass and twist corrections must be taken into account
\cite{DIS_SF}. The choice which is adopted in most MC codes is to apply structure functions with
corrections modelled in analogy to the electron scattering case. Corrections which are available in
the literature are applied to LO GRV98 PDF's \cite{BY}. Their form is closely related to the issue
of quark-hadron duality: the DIS structure functions describe on average the electron scattering
data in the resonance region \cite{Duality}. It is an open problem whether quark-hadron duality
should hold also in $\nu$N scattering. There has been recently a lot of investigation in this field
\cite{Nu_duality}. The theoretical analysis of the duality has been done in the framework of
$SU(6)$ quark model of resonances \cite{Close}. It is not clear whether arguments valid for the
vector part of the hadronic current should hold true also for its axial counterpart. Another
problem is to understand what happens in the kinematical region $Q^2<0.5$~GeV$^2$, where one does
not expect the quark-hadron duality to be present. In electron scattering it is known that
$F_2^{eN}\rightarrow Q^2$ and $F_L^{eN}\rightarrow Q^4$, but for neutrino structure functions the
presence of the axial current, which is not conserved makes the situation more complicated
\cite{Bodek_RS}.

Once the inclusive cross section is calculated one has to evaluate contributions from exclusive
channels. One possibility is to use the KNO theory which provides average multiplicities of
particles in the final state \cite{Gall}. The only remaining problem is then to redistribute to the
particles energy and momentum transfer. Another strategy is to use the LUND fragmentation and
hadronization routines \cite{Lund,Wroclaw}.

In MC implementation of either scheme it is necessary to decide in
several important points. Where should be a boundary between DIS
and resonance (SPP) contributions? The very definition of the RS
model suggests that one should define $W_{cut}=2$~GeV. Some
authors argue that the RS model underestimates the cross section
at higher $W$ and the better choice is $W_{cut}\sim 1.7$~GeV
\cite{Bodek_RS}. This is the choice adopted by the authors of the
NEUGEN/GENIE MC code \cite{Gall}. The comprehensive analysis of
all the available date led other authors to the conclusion that
one should take $W_{cut}\sim 1.5$~GeV \cite{Vadim}. This is
approximately the choice implemented in the WROCLAW MC generator
\cite{Wroclaw} where only $\Delta$ resonance contribution is
included.

\section{Nuclear effects - generalities}

The treatment of nuclear effects depends on the neutrino energy. In the few GeV energy region one
can rely on the picture in which neutrino interacts with individual nucleons inside nucleus
(impulse approximation (IA)). It is not completely clear starting from which energies IA is the
correct approach. Some authors argue that the Fermi gas model (the simplest mean field theory
realization of IA) works well for electron neutrino energies $E_\nu\geq 200$~MeV \cite{Vogel}.
Other authors are more conservative and argue that the IA picture makes sense for momentum
transfers $q\geq 400$~MeV, which translates into higher neutrino energies \cite{Co}. For example,
for neutrino energy $E_\nu=0.8$~GeV the contribution to the quasi-elastic cross section from the
momentum transfers $q<400$~MeV is $\sim 20\%$.

The simplest realization of the IA is known as PWIA (plane wave
impulse approximation): one assumes that the nucleon produced in
the primary vertex leaves nucleus without further re-interactions.
This is an obvious oversimplification and it is better to include
FSI (final state interactions) effects. A possible systematic
approach to deal with FSI is known as DWIA (distorted wave impulse
approximation) \cite{Maieron}. In MC codes FSI effects are usually
treated by means of inter-nuclear cascade modules \cite{FSI_MC}.
The propagation of nucleons, pions and other particles inside
nucleus is semiclassical. It is important to implement the concept
of the formation zone. Many other theoretical schemas to deal with
FSI has been developed as well \cite{FSI_th}.

\section{Nuclear effects - some models}

The advantage of the Fermi gas (FG) model is that it is easily
applicable in MC routines. The basic FG model is defined by just
two parameters: Fermi momentum and binding energy \cite{SM}. It is
necessary to deal with the problem of how to calculate off-shell
nucleon matrix elements and the de~Forest prescription is the
common way to handle it \cite{dF}. One has also to decide about
the kinematics. Smith-Moniz approach is the simplest choice and
the other is to take into account the recoil nucleus momentum
\cite{BR}. An improvement to the above versions of the FG model is
obtained in the framework of LDA (local density approximation):
the Fermi momentum becomes a local quantity according to the
density profile of the nucleus \cite{LDA}. Further improvement is
introduced by modifying the momentum distribution of the nucleons
by adding the high momentum tail \cite{BR}. In the framework of
the FG model the only FSI effect is Pauli blocking.

%%%%%%%%%%%%%%%%%%%%%%%%%%%%%%%%%%%%%%%%%%%%%%%%%%%%%%%%%%%%%%%%%%%
\begin{figure}
\begin{tabular}{c c}
    \includegraphics[scale=0.4]{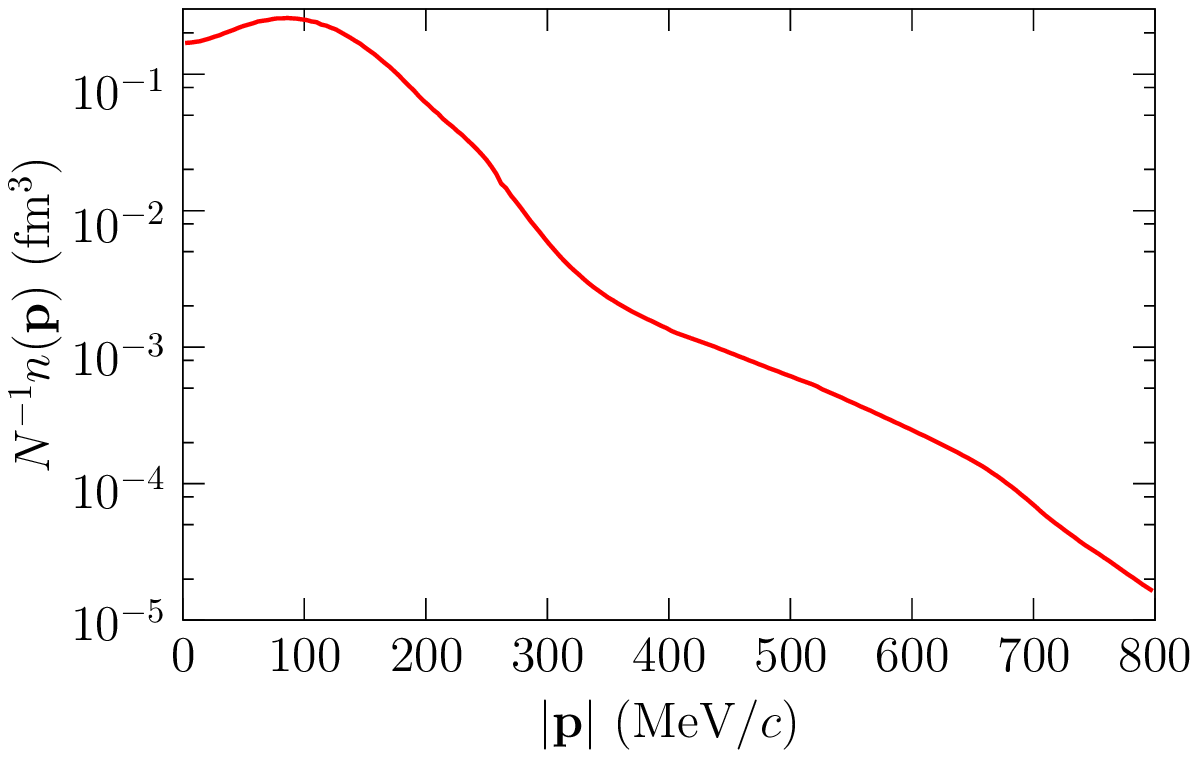}  \includegraphics[scale=0.4]{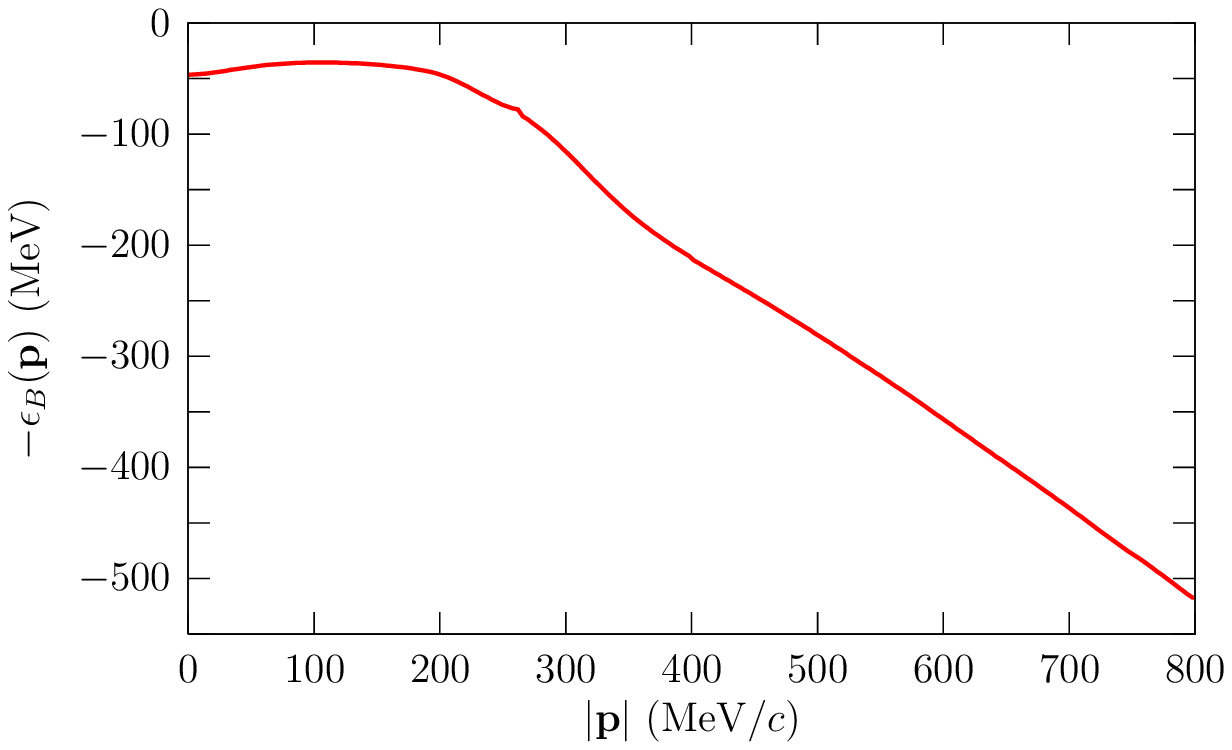}\\
    \includegraphics[scale=0.8]{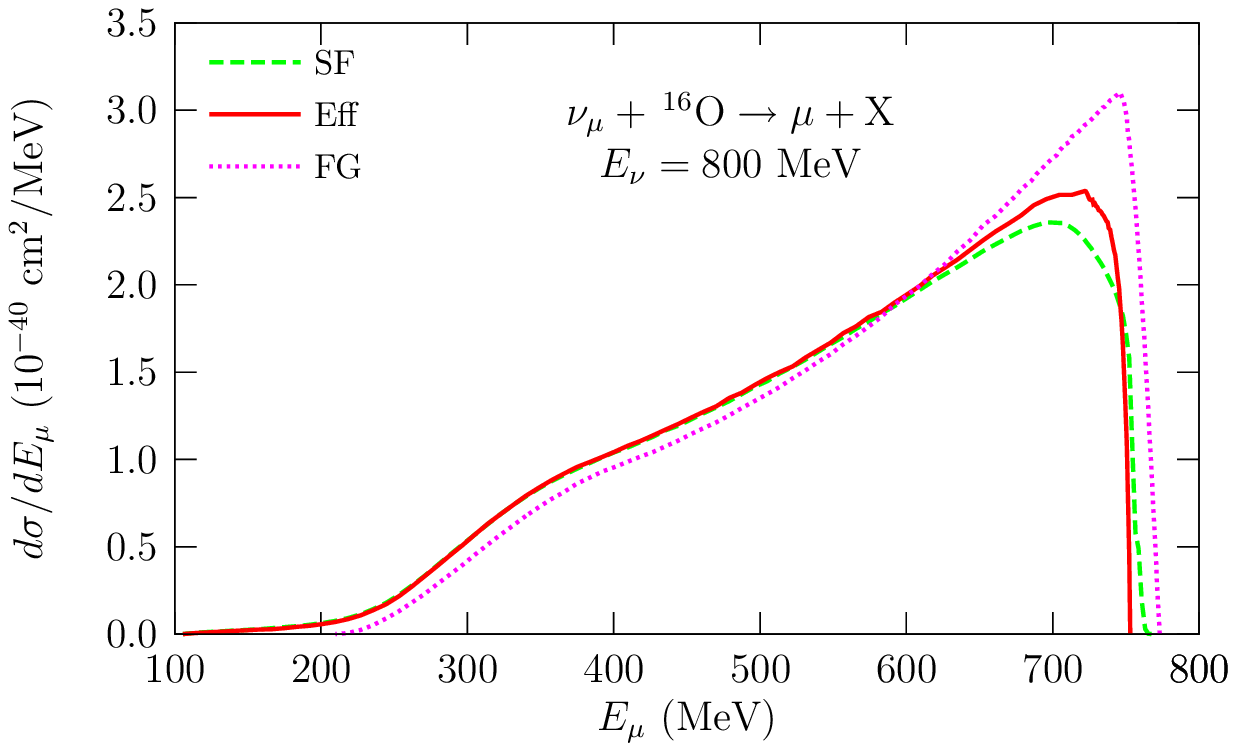}\\
\end{tabular}
  \caption{The {\it effective} description \cite{AS} based on
O. Benhar's oxygen spectral function \cite{Omar_oxygen}.
  The upper two plots show the momentum distribution (left-hand side) and the
average momentum dependent binding energy (right-hand side). Below a comparison is shown between
three modelings of nuclear effects: by Fermi gas (FG), spectral function (SF) and the {\it
effective} one (Eff).}
\end{figure}

%%%%%%%%%%%%%%%%%%%%%%%%%%%%%%%%%%%%%%%%%%%%%%%%%%%%%%%%%%%%%%%%%%

The spectral function (SF) approach represents an improvement with
respect to FG model by providing a realistic probability
distribution of momenta and binding energies of nucleons inside
nuclei \cite{Omar}. Theoretical models of SF are obtained by
combining the mean field (shell model) and the correlated part.
The second one is relevant at higher values of momenta and binding
energies and comes from short range interactions due to correlated
high momentum pairs of nucleons \cite{Cioffi}. Recently the
correlated part of the SF has been directly measured in electron
experiments \cite{Rohe}.

Reliable models of SF exist for lighter nuclei up to oxygen
\cite{Omar_oxygen}. The mean field part of SF is clearly seen as
probability distribution peaks at $E\sim -42$~MeV ($1s$), $E\sim
-19$~MeV ($1p_{3/2}$) and $E\sim -13$~MeV ($1p_{1/2}$). The
characteristic feature of the $1s$ level is that it is smeared out
in $E$.

MC implementation of SF approach is straightforward. One can also
use the {\it effective} approach in which the relevant information
about the spectral function is contained in two functions: the
probability distribution of nucleons momenta and the average
momentum dependent binding energy \cite{AS}.

In order to construct SF for heavier nuclei it is necessary to
know the energy levels and spectroscopic factors. The correlated
part of SF is universal and depends only on the nucleus size
\cite{Omar}.

SF can be also used to model SPP in the resonance region. A new
issue is the dependence of the $\Delta$ resonance width on the
nuclear matter \cite{Delta_nucleus}.

In the context of the DIS formalism there are specific methods to deal with nuclear effects to
describe the shadowing, anti-shadowing {\it etc}. Recently a comprehensive model has been proposed
to describe all the effects in the unique theoretical frame \cite{KP}. It includes also Fermi
motion effects and in combining it with nuclear effects for quasi-elastic and SPP one has to be
careful to avoid double counting.

A step beyond IA would be to include contributions from 2-body
current. Is is believed they are necessary in order to explain the
excess of the cross section in the DIP region between
quasi-elastic and $\Delta$ excitation peaks \cite{DIP}. The
problem with 2-body currents is that the computations which must
be performed are algebraically very involved \cite{Wanda}. Some
authors tried to approximate the 2-body contribution to the
neutrino cross section with the conclusion that it is very
important: for $E_\nu = 700$~MeV it is responsible for up to $\sim
25\%$ of the cross section in the kinematical region of energy
transfer $\omega\in (80, 250)$~MeV \cite{Marteau}.

\section{Final remarks}

There has been efforts in the past to create a universal MC code
to describe neutrino interactions \cite{NuInt}. So far each
experiment uses its own MC focused on particular neutrino energy
spectrum, target, detection techniques etc. The most promising
ongoing project to construct a universal MC is that of GENIE
\cite{Genie}.

All the theoretical considerations is this review were subject to
big experimental uncertainty in $\nu$N cross sections. The good
news are that the MINERvA experiment is under way \cite{Minerva}.
It will enable us to settle a lot of unknowns in free nucleon  and
nuclei targets cross sections.\\
\\
{\bf Acknowledgments}\\
\\
The authors were partially supported by the Polish State Committee for Scientific Research (KBN),
grant No. 105/E-344/SPB/ICARUS/P-03/DZ211/2003-2005. The authors thank Cezary Juszczak, Artur
Ankowski and Krzysztof Graczyk for friendly collaboration and many fruitful discussions.

\end{document}